\documentclass[twocolumn,showpacs,preprintnumbers,amsmath,amssymb]{revtex4}

\usepackage{graphicx}
\usepackage{dcolumn}
\usepackage{bm}

\begin{document}

\title{
Dipolar coupling and multidomain states 
in perpendicularly polarized nanostructures}

\author{Alexei N.\ Bogdanov}
\altaffiliation[Permanent address: ]%
{Donetsk Institute for Physics and Technology,
340114 Donetsk, Ukraine
}
\email{a.bogdanov@ifw-dresden.de}
\author{Ulrich K.\ R\"o\ss ler}
\thanks
{Corresponding author 
}
\email{u.roessler@ifw-dresden.de}

\affiliation{
IFW Dresden,\\
P.O. Box 270116, D--01171 Dresden, Germany
}%

\date{\today}

\begin{abstract}
{
Stripe states in multilayer systems 
with perpendicular polarization are investigated
by analytical calculations within a general 
continuum approach, applicable to ferromagnetic, 
ferroelectric, or ferroelastic nanoscale superlattices. 
The competition between 
the long-range depolarization effect and 
short-range interlayer couplings 
can stabilize monodomain states and unusual stripe phase 
ground states with antiparallel polarization in adjacent layers.
Geometric parameters of stable stripe domain states 
and the phase transitions lines 
between single domain states, aligned 
and antialigned stripe states have been derived.
The theory is applied to analyze multidomain states
and phase transitions in antiferromagnetically
coupled multilayers [CoPt]/Ru.
}
\end{abstract}


         
\maketitle


%
Multidomain states considerably influence
physical properties of condense matter
systems with spontaneous polarization.
Such spatially inhomogeneous patterns
form ground states of
 ferromagnetic \cite{Kittel,Hubert98}, 
ferroelectric \cite{Scott06}, 
or ferroelastic \cite{Bratkovsky01} films.
Recently multidomain structures  have been observed
in nanoscale magnetic films and multilayers
with  strong perpendicular anisotropy 
\cite{Gehanno97,Hamada02,Hellwig03,Itoh03,PRB04}
and in ferroelectric superlattices 
\cite{Streiffer02,Scott06,Lee04}.
Similar spatially modulated states can also  arise
in polar or magnetic liquid crystals 
\cite{Lagerwall,Sayar03}, 
polar multiblock copolymer layers \cite{Ruzette05},
in superconducting films or 
magnetic-superconductor hybrids \cite{Gillijns05},
and in shape memory alloy films \cite{Dong04}.
Multilayer systems with perpendicular polarization components
provide ideal experimental models to investigate fundamental 
aspects of ordered structures and stable pattern formation 
in confining geometries.  
Control of such regular depolarization patterns is also
of practical interest. 
E.g., patterns may provide templates, 
which could be decorated by nanoparticles or macromolecules, 
or they may be used to calibrate imaging techniques.
Particular interest in ferroelectric superlattices
is driven by the exciting possibility of
using ferroelectric nanostructures
in nonvolatile memory applications, 
new microelectromechanical systems (MEMS), 
and for nonlinear optics devices \cite{Scott06}. 
In nanomagnetism, antiferromagnetically coupled
superlattices with strong perpendicular 
anisotropy \cite{Hamada02,Hellwig03,Itoh03}
are considered as promising candidates 
for nonvolatile magnetic recording media 
and other applications \cite{Fullerton03}.
%
%
According to recent experiments
\cite{Hamada02,Hellwig03,Scott06}
these nanoscale superlattices 
are characterized by new multidomain
states, unusual depolarization processes,
and other specific effects which
have no counterpart in other classes
of media with perpendicular polarization.

We present here a detailed analysis of
multidomain states
in magnetic nanoscale superlattices
with perpendicular polarization components.
We derive simplified micromagnetic equations 
for equilibrium parameters of stripe domains
in nanoscale multilayer systems.
%
%
These mathematical tools provide
a clear description of the
multidomain processes and
reveal the physical mechanism underlying 
their unconventional properties.
It was found that in contrast to 
other bulk and nanomagnetic systems, 
the magnetic states here are determined 
by a close competition between interlayer 
exchange and dipolar couplings. 
The enhanced stray field couplings are responsible
for the unusual switching processes and 
specific transformation of the domain patterns observed
in synthetic metamagnetic multilayer systems, as [CoPt]/Ru
and others \cite{Hamada02,Hellwig03,Itoh03}.
The depolarization effects
revealed in this paper
have a universal character.
They arise similarly in ferroelectric
superlattices and in other nanosized polarized media.

\begin{figure}
%
\includegraphics[width=7.0cm]{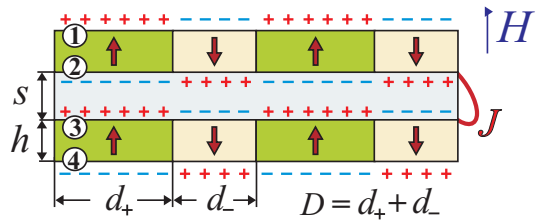}
\caption{
\label{StripeGeometry}
A fragment of a superlattice:
two ferromagnetic
nanolayers (of thickness $h$) 
with stripe domains are
separated by a nonmagnetic
spacer of thickness $s$.
The domains are coupled
by the exchange ($J$) 
and magnetostatic forces.
}
\end{figure}

For definiteness we consider stripe domains in
magnetic nanolayers.
We analyse superlattices 
consisting of $N$ identical layers
with thickness $h$ separated by spacer layers
of thickness $s$ (Fig.~\ref{StripeGeometry}).
The perpendicular anisotropy fixes the easy 
magnetization direction.
Within the single layers the magnetization 
${\mathbf{ M}_i}(\mathbf{r})$, 
may be spatially inhomogeneous. 
The energy can be written in a phenomenological approach as 
\begin{eqnarray}
\label{energy0}
& & W_N  =  \sum_{i=1}^{N-1} \int  \int j(\mathbf{r}_i,\mathbf{r}_j)
\, \mathbf{M}_{i} \cdot
\mathbf{M}_{i+1} dv_i\,dv_j +
\\
& & \sum_{i=1}^{N} \int \left[-\frac{K}{2}\left(\mathbf{M}_i 
\cdot\mathbf{n} \right)^2
 -\mathbf{H}^{(e)}\cdot \mathbf{M}_i
-\frac{1}{2}\mathbf{H}_d \cdot \mathbf{M}_i \right] d v_i\,
\nonumber
\end{eqnarray}
where integrals are over the volume $v_i$ of the single layers.
$\mathbf{H}^{(e)}$ and $\mathbf{H}_d (\mathbf{r})$
are the externally applied 
and the depolarizing magnetic fields,
respectively.
The unity vector ${{\mathbf n}}$ designates 
the normal to the film. 
$K > 0$ is a perpendicular anisotropy.
The short-range interlayer coupling parameter
between (nearly) homogeneous layers 
$J = \int\int  j(\mathbf{r}_i,\mathbf{r}_j) dv_i\,dv_j  > (<)$~0 
favours (anti)parallel orientation of the neighbouring layers. 
To investigate general effects of competing 
stray field and exchange interlayer interactions,
we consider a simple model of a multidomain structure,
namely stripe domains with antiparallel magnetization 
of magnitude $M \equiv |\mathbf{M}_i=\textrm{const}$ 
within all the magnetic layers. 
The adjacent domains are
separated by \textit{thin} domain 
walls with energy density $\sigma$.
The magnetic field $\mathbf{H}$ is applied
perpendicular to the layer surfaces.
The stripe structure is 
described 
by the widths of domains $d_{\pm}$
that are 
polarized 
in the directions parallel ($+$) and antiparallel ($-$)
to the field.
The geometry of stripes in a multilayer 
structure is conveniently defined by
the stripe period $D  = d_{+}+d_{-}$, 
and a set of reduced parameters
\begin{eqnarray}
\label{pq}
q = \frac{d_{+}-d_{-}}{D}, \; 
p = 2\pi\frac{h}{D},  \quad
\nu  =  \frac{s}{h}, \; 
\tau  = 1+\nu\,, 
\end{eqnarray}
where $q$ is proportional to the average magnetization
in a stripe structure, $p$ is the reduced 
thickness of single ferromagnetic layers, 
$\nu$ is the thickness ratio between 
magnetized layer and interlayer,
and $\tau$ fixes the superlattice period
(Fig. \ref{StripeGeometry}).

The reduced energy $w_N = W_N/(2 \pi M^2 N$)
of a system with $N$ can be written
\begin{eqnarray}
\label{energy1}
 w_N = \frac{2 \Lambda p}{\pi^2} - \frac{H q}{2 \pi M}
+ \varkappa \left( 1 - \frac{1}{N} \right) + w_m (p,q)\,.
\end{eqnarray}
In Eq.~(\ref{energy1}), the term linear in $p$ 
is the energy of the domain walls. 
The next two terms are Zeeman energy and
the short-range interlayer coupling, respectively. 
The stray field energy $w_m (p,q)$
must be derived by solving
the corresponding magnetostatic problem
\cite{Suna85}.
The reduced energy (\ref{energy1})
depends on the two dimensionless materials parameters
\begin{eqnarray}
\label{kappa}
\varkappa = \frac{J}{2\pi M^2}, \quad
\Lambda =  \pi \frac{l}{h}\,. 
\end{eqnarray}
The strengths of the exchange coupling 
is measured by $\varkappa $  given by 
the ratio between the exchange and stray field energies.
The parameter $\Lambda$ characterizes 
the balance between the domain wall energy and 
stray field energies. 
It is fixed by the ratio of 
\textit{characteristic  length}, $l=\sigma/(4\pi\, M^2)$,
which is a fundamental material parameter,
see Ref.~\onlinecite{Hubert98}) and the 
thickness of the magnetic layers.

\begin{figure}
%
\includegraphics[width=7.0cm]{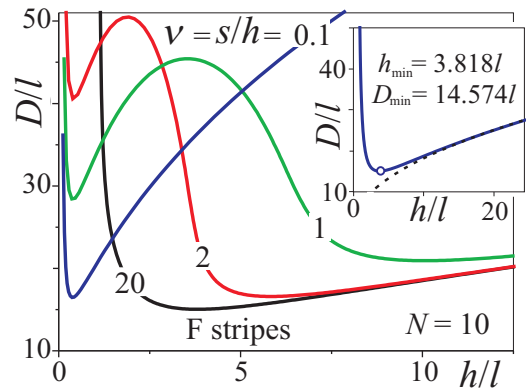}
\caption{
\label{FMPeriods}
The equilibrium reduced period $D/l$ as a function
of the reduced layer thickness $h/l$ for ferro (F)
stripes in an $N=10$ multilayer for different values 
of parameter $\nu$.
Inset shows the function $D/l$ ($h/l$)
for an isolated layer \cite{MalekKambersky}
and the parameters of the minimum point.
Dashed line indicates $D \propto \sqrt{h}$
fit corresponding to Kittel theory.
}
\end{figure}

\begin{figure}
%
\includegraphics[width=7.5cm]{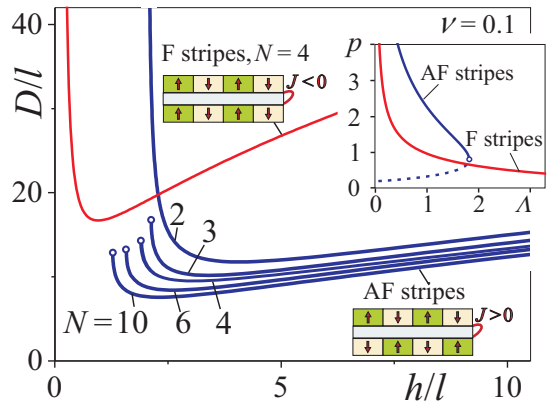}
\caption{
\label{SolutionsD}
The solutions $D/l$ ($h/l$) for ferro and
antiferro (AF) stripes in 
systems with $\nu = 0.1$.
AF stripes exist for thickness 
larger than the critical thickness $h_{cr}$. 
For a two-layer systems ($N =2$) the
stripe period goes to infinity at the critical thickness.
For systems with $N > 2$ the critical 
stripe period $D_{cr}$ has finite values.
The thin (red) line shows 
the dependence $D(h)/l$ for the period of 
F stripes in the case $N=4$. 
The corresponding solutions $p(\Lambda)$
of Eq. (\ref{equationP}) for these parameters
are plotted in the Inset: F stripes (red) and
AF (blue) lines. Dashed line indicates
unstable solutions.
}
\end{figure}

Due to the mathematical identity of
electro- and magnetostatic equations
\cite{Hubert98} the multilayer
with stripes can be thought of
as  a set of  planes with ``charged''
stripes (Fig. \ref{StripeGeometry}).
For two such planes separated by an interlayer with 
thickness $a=\omega h$,
the magnetostatic energy is
\begin{eqnarray}
\label{energyMS1}
f(\omega)  = \frac{4}{p}
\sum_{n=1}^{\infty} 
\frac{1 -(-1)^n \cos (nq)}{n^3} \exp(-n\,p\,\omega)\,.
\end{eqnarray}
The stray-field energy of the ``charges''
within the same plane is
$w_m^{(0)} = q^2 +f(0)$.
The dipolar coupling energy
between the ``poles'' on different
sides of the same layer is $f(1)$.
Hence, the magnetostatic energy of
individual layers is \cite{MalekKambersky,KooyEnz60}
\begin{eqnarray}
w_m^{(\textrm{self})} =q^2 +f(0)-f(1)
\label{energyMS0}
\end{eqnarray}
The stray field energy of the multilayer,
$w_m$ 
originally derived by Suna,\cite{Suna85}
can be written as a sum of 
``self'' energies of the layers and 
interactions between them 
\begin{eqnarray}
\label{energyMS}
w_m & = &  w_m^{(\textrm{self})}+ w_m^{(\textrm{int})}, 
\\
w_m^{(\textrm{int})} & = & 
\pm  \frac{1}{N} \sum_{k=1}^{N-1} (N-k)F_k\,,
\nonumber\\
F_k &  = &  f(\tau k +1) + f(\tau k -1) -2 f(\tau k)\,.
\nonumber
\end{eqnarray}
The upper (lower) signs in $w_m^{(\textrm{int})}$
correspond to parallel (antiparallel) arrangement
of the polarization in adjacent layers
(Figs.~\ref{FMPeriods},~\ref{SolutionsD}).
We denote these modes 
as \textit{ferro} (F) or \textit{antiferro} 
(AF) stripes. 
The factors $F_k$ equals
stray field coupling energies 
between two layers separated by distance $\tau k$. 
They are composed of the four contributions
of the magnetodipole coupling 
between pairs of the planes bounding 
the layers. 
In particular, for two adjacent
layers, $k =1$ (Fig. \ref{StripeGeometry}),
the interactions  
$1\leftrightarrow 3$ 
and $2\leftrightarrow 4$ yield
equal (positive) energy contributions $f(\tau)$,
while 
$1\leftrightarrow 4$ and $2\leftrightarrow 3$
yield negative energy contributions,
$f(\tau+1)$ and $f(\tau-1)$, correspondingly.

In common polarized systems 
with characteristic sizes  far
beyond the nanoscale range
the equilibrium domain sizes are usually
much smaller than the individual 
layer thicknesses, $p \gg 1$.
Numerous observations indicate
that, as soon as domain sizes approaches
the layer thickness, coercitivity suppresses
the formation of regular multidomain patterns
\cite{Hubert98}).
This establishes a natural limit for domain
sizes in classical systems.
For $D \ll h$ there is no effective 
dipole interaction  between different surfaces.
Hence, in Eq.(\ref{energyMS})
for all nonzero $\omega$, one hase $ f(\omega) \ll f(0)$,
and the stray field energy  of the multilayer
is reduced to $w_m^{(0)}$ \cite{Kittel}.
In such  decoupled superlattices the multidomain
states should have similar properties as those
in isolated layers.

On the contrary, in perpendicular polarized 
nanoscale films and multilayers
the periods of \textit{regular}
multidomain patterns is of the same
order as their thicknesses
or exceed these thicknesses
\cite{Gehanno97,Hellwig03,Streiffer02}.
In such systems dipole interactions 
between different surfaces have
a sizable effect.
Mathematically, this is seen from slowly converging 
sums of interaction terms between poles 
far apart and on different internal surfaces. 
For such structures, numerical evaluation becomes arduous, 
and sharpened analytical methods are required.
To overcome the slow convergence in $w_m$,
we extend the method introduced in \cite{FTT80}. 
With the help of the identity 
$ \int_0^{\infty} t^{(m-1)}\exp(-nt) dt = m!/n^m$,
the infinite sums in Eq. (\ref{energyMS1})
can be transformed into integrals on the interval
$[0,1]$

\begin{eqnarray}
f(\omega) - f(0)= \pi^2 (1-q^2) -2p \Omega(\omega)
\label{int1}
\end{eqnarray}

where
\begin{eqnarray}
\label{Omega}
\Omega (\omega)= \omega^2 \int_0^1 (1-t)\ln 
\left[1 + \frac{ \cos^2 \left(\pi q/2 \right)}
{ \sinh^2 \left(\omega p t/2 \right)}\right]d t\,.
\end{eqnarray}

Then
the dipolar stray field 
energy (\ref{energyMS})
can be written as 
\begin{eqnarray}
\label{energyStripeMC}
w_m =  1- \frac{2 p}{\pi^2} \Omega (1) \pm
\frac{2 p }{\pi^2 N} \sum_{k=1}^{N-1} (N-k)\Xi_k (\tau k),
\end{eqnarray}

\begin{eqnarray}
\label{energyStripeMC2}
\Xi_k (\tau k) = 2 \Omega(\tau k)
-\Omega(\tau k +1) 
-\Omega(\tau k -1)\,.
\end{eqnarray}
Minimization of $w_N$ (\ref{energy1}) with respect 
to $p$ and $q$ yields the equilibrium parameters
of the stripes in the multilayers.
By the form of the multilayer energy Eq.~(\ref{energy1})
the widths of the stripes are independent of the 
interlayer couplings $\varkappa$.
A complete analysis of these equations in 
applied field will be published elsewhere 
\cite{PRBstripes}.
Here we investigate the ground state of the
system in zero applied field, $\mathbf{H} = 0$.
In this case  $q =0$ and the parameter
$p$ is derived from the equation
$d w_N /d p = 0$
\begin{eqnarray}
\Lambda = \Omega_p (1)
\mp \frac{1}{N}
\sum_{k=1}^{N-1} (N-k)\Xi_k^{(p)} (\tau k +1)
\label{equationP})
\end{eqnarray}
where $\Xi_k^{(p)} (\tau k +1) = 
2 \Omega_p (\tau k)
-\Omega_p (\tau k +1) 
-\Omega_p (\tau k -1)$, and
$\Omega_p (\omega) = d \Omega /d p(\omega)$. 
After elementary transformations 
this function can be written 
\begin{eqnarray}
\Omega_p (\omega)= -2 \omega^2 \int_0^1 t 
\ln \left[ \tanh \left( \frac{p t}{2} \right) \right] d t\,.
\label{omegaP}
\end{eqnarray}

Typical solutions of Eq. (\ref{equationP})
are presented in Figs. \ref{FMPeriods} 
and \ref{SolutionsD}.

 \textit{Ferro stripes}. These solutions 
exist for any  layer thickness Fig.~\ref{FMPeriods}. 
In the limit of thick layers the period $D$ increases,
while it tends to infinity as $h$ tends to zero.
The role of the stray-field couplings changes
with the ratio of the interlayer thickness to 
the thickness of the polarized layers, $\nu$.
For the limiting cases of small 
and large 
thickness of the spacer layer
given by the limits $\nu\ll 1$ or $\nu \gg 1$, respectively,
the dependence $D(h)$ approaches 
a behaviour of isolated layers with 
thickness $hN$ and $h$ correspondingly.
In the limit of small ratios $\nu$ 
the stripe period $D(h)$ approaches the period 
for a F stripe state in a single layer with an effective 
total polarized layer thickness $hN$. 
For very large separation between the polarized layers, 
i.e., for large ratios $\nu$ the stripe period is determined 
by the properties of the decoupled single layers with thickness $h$.
The limiting solution for an isolated layer $D(h)$
(Inset Fig. \ref{FMPeriods})
obtained by \cite{MalekKambersky} 
(see also Ref.~\onlinecite{Hubert98}) has 
a minimum point with parameters
($h_{min}/l = 0.96067$, $D_{min}/l = 16.3136$). 
The nonmonotonic behaviour of $D(h)$
reflects the antagonistic role of magnetic charges
in the formation of the equilibrium stripes.
In the case of small domains  $D \ll h$, which
is typical for classical systems, 
the dipole interaction between different 
surfaces of the layer is negligibly small, 
and only the interaction between charges 
on the same surface give a contribution 
to the stray field energy.
For stripes with sizes $D \geq h$ the
interaction between charges from different
surfaces becomes a noticeable effect
and counteracts the interactions between
charges on the same surface.
This can be understood as a screening effect.
As the layer thickness decreases 
this screening effect becomes stronger 
and suppresses the stray field energy.
As a result, for $h < h_{min}$ the extension
of domains decreases both the sum of domain wall  energies
and the stray field energy, and the domain period 
increases exponentially with decreasing 
layer thickness.
These simple energetical arguments demonstrate
that in nanoscale multilayers, where domain sizes
usually considerably exceed the thicknesses of 
magnetic and interlayer, $h$ and $s$,
the interaction between magnetic poles through
the stack strongly influences the equilibrium
magnetic states.
%
%
Due to this effect
a non-monotonous dependence of stripe periods $D(h)$ arises 
in multilayers between the limiting cases of large/small values 
both of $\nu$ and $h$, as shown in \ref{FMPeriods}. 

\textit{AF stripes}. In this case the solutions
of Eq. (\ref{equationP})  $p(\Lambda)$ exist 
on the finite interval $\Lambda < \Lambda_{cr}$
and consist of two branches with stable and unstable
solutions (Fig. \ref{SolutionsD}, Inset).
Correspondingly the equilibrium stripe states 
exist only in a finite 
range of the thickness $h < h_{cr}$ (Fig. \ref{SolutionsD}).
At a critical thickness $h_{cr}$ the period
reaches a critical value $D_{cr}$ 
for multilayer systems with $N > 2$ and tends to infinity
for two-layer system, $N=2$, (Fig. \ref{SolutionsD}).

\begin{figure}
%
\includegraphics[width=7.5cm]{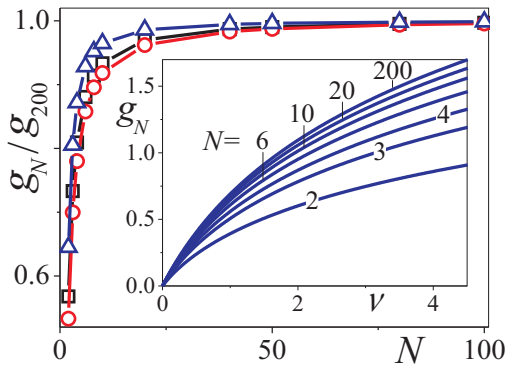}
\caption{
\label{gN}
The reduced values of the factor $g_N$ as
a function of $N$ and different values
of $\nu$: 1 ($\triangle$), 5 ($\square$), 0.1 ($\bigcirc$).
Inset gives functions $g_N (\nu)$
for different values $N$.
}
\end{figure}

\begin{figure}
\includegraphics[width=7.5cm]{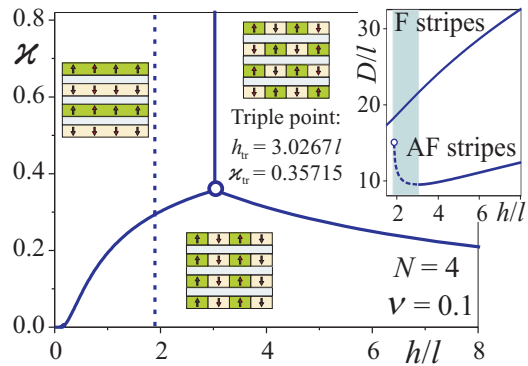}
\caption{
\label{Kappah}
The phase diagram of magnetic states in variables
$h/l$ and the reduced coupling $\varkappa$ favouring
antiparallel orientation of adjacent layers for $N=4$.
Thick lines indicate the first-order transitions
between different phases. 
They meet in a triple point.
The dashed line shows the stability limit
of the AF stripes.
Inset shows a difference between stripe periods
in F and AF modes.
Between the triple point 
and the critical point AF domains
are metastable.
}
\end{figure}

\begin{figure}
\includegraphics[width=8.00cm]{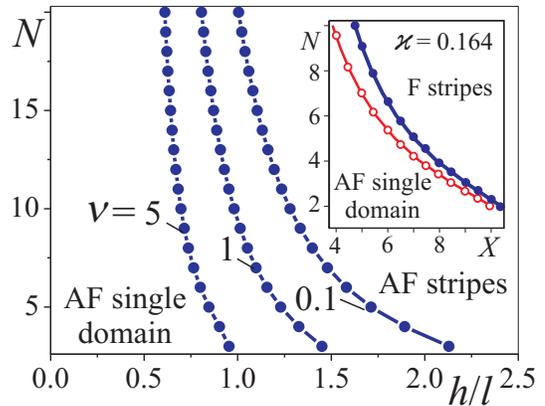}
\caption{
\label{hN}
The phase diagram in variables
$h/l$  and $N$ with the
critical lines between the antiferro
single domain and the multidomain states.
The transitional line
between F stripes 
and AF single domain state (Inset, solid points) is in a close agreement
with results obtained for magnetic Co/Pt multilayers
in \cite{Hellwig03}.
The line with hollow points indicates the transition
in a model with zero internal charges ($g_N =0$).
}
\end{figure}

In multilayers with a ferromagnetic coupling ($J < 0$) both 
the stray fields and exchange interactions favour parallel
arrangement of the magnetization across the stack.
Hence, the F-stripe mode is the ground state in such multilayers.
On the contrary, in systems with
antiferromagnetic interlayer couplings, $J> 0$,
the dipole and exchange forces have competing character. 
As a result, three different equilibrium phases can exist in
such multilayers  in zero field,
see phase diagram Fig.~\ref{Kappah}.
For sufficiently strong interlayer couplings
a monodomain phase with antiparallel correlations of adjacent layers 
across the stack, and an AF stripe phase exist. 
The transition between the monodomain and AF stripe states 
for the two-layer system, $N=2$, is continuous, 
as seen from the divergence of $D(h)$ (Fig.~\ref{SolutionsD}).
For systems with $N \geq 3$ the transition is first-order.
For weaker interlayer coupling $\varkappa$, 
the F stripe phase becomes stable. 
The transition between the F stripes and 
the AF stripes or monodomain state is a topological
transition, and it is always first-order.
At a triple point, that weakly depends on 
the multilayer repetitions $N$,
all three phases coexist.
Lines of first order transitions between the AF monodomain and 
AF stripe phase in dependence on $N$ and layer thickness $h$ 
are shown in Fig.~\ref{hN}.

In the limit of large domains ($D \geq l$)
the expansion of the integral (\ref{Omega}) 
for $q=0$ yields
$\Omega (\omega) = 3/2 - \ln (p \omega/2)$
and energy (\ref{energy1}) can be transformed
in the form
\begin{eqnarray}
 w_N  =   1 - \frac{2 p_N}{\pi^2} 
\left[ \frac{3}{2} -\ln \left( \frac{p_N}{2} \right) 
-\tilde {\Lambda}_N \right]  
 + \varkappa \left( 1 - \frac{1}{N} \right),
\label{energy3}
\end{eqnarray}
where 
$p_N =pN= 2 \pi (hN)/D$, 
$ \tilde {\Lambda}_N = \Lambda/N + g_N(\nu)$.
Then, the energy $w_N$ (\ref{energy3}) has the 
form of the energy for an \textit{isolated} layer
(see \cite{FTT80}) with ``effective'' thickness $hN$.
This means that in a system with large
domains ($D \gg h$)
the multilayer stack behaves as an effectively
coupled single layer and the domain period
approaches the solution for a single layer 
with a total polarized thickness $hN$
with an effective characteristic parameter 
$\tilde {\Lambda}$.

The function $g_N(\nu)$ describes the influence
of finite spacers on the magnetic
properties of the multilayer
via the redistribution of the internal ``charges''
within the stack.
This function is given by
\begin{eqnarray}
 g_N  = - \frac{1}{N^2} \sum_{k=1}^{N-1} (N-k)\widetilde{G}(\nu)
 -\ln N\,, 
\label{gN}
\end{eqnarray}
where 
$\widetilde{G}(\nu) = 2\tilde{g}(\tau k)-\tilde{g} (\tau k +1) 
- \tilde{g}  (\tau k -1) $
and
$\tilde{g}(\omega) = \omega^2 \ln(\omega)$.
The dependence of $g_N(\nu)$ on the number of layers $N$ 
and the ratio $\nu=s/h$ is shown in Fig.~\ref{gN}.
In the limit of small $\nu$ 
this function behaves as $g_N = a_N \nu$ 
where the factor 
\begin{eqnarray}
 a_N  = 2 (1 -1/N ) \ln N 
- 4 N^{-2} \sum_{k=1}^{N-1} k \ln k 
\label{aN}
\end{eqnarray}
varies from $\ln(2)$ for $N =2$ to 
unity as $N$ tends to infinity.
The equilibrium domain period in the limit of large $N$
is (cf.~\cite{FTT80})
\begin{eqnarray}
 D = \pi hN \exp(\tilde {\Lambda}_N-1/2)\,.
\label{Dn}
\end{eqnarray}

Recently, stripe domains have been investigated 
in Co/Pt multilayers
\cite{Hamada02,Hellwig03,Hellwig05}.
AF stripes have been observed in Co/Pt multilayers \cite{Hamada02}.
F stripes have been investigated in a set of multilayers
[Co(4\AA) Pt (7\AA)]$_X$  (with $X$ from 5 to 160)
\cite{Hellwig05}.
The average domain width $D/2$ in our notation
plotted versus the total multilayer  thickness is close
to the line typical for an isolated layer 
(Inset, Fig.~\ref{FMPeriods}), 
the minimum point $(h_{min},D_{min})$
is in the region $X \approx 20$ with a  period $D$  
about 150~nm. 
This system has $\nu_f$ = 7/4=1.75.
However, the existing data 
are insufficient for detailed analysis 
and comparison with a theoretical dependencies 
for $D(h)$ in the Fig.~\ref{FMPeriods}.
More substantial results have been obtained on the investigation
of antiferromagnetically coupled (via Ru) ferromagnetic blocks
[[Co(4\AA) Pt (7\AA)]$_{X-1}$Co(4\AA) Ru(9\AA)]$_N$
(with  $N = 2$ to $10$  and $X= 2 $ to $12$), Ref.~[\onlinecite{Hellwig03}].
Strictly speaking magnetic properties of
such ferromagnetic  multilayers may 
strongly differ from those of single
layers (see Fig. \ref{FMPeriods}).
However, according to the results of \cite{Hellwig05} 
these blocks can be modelled by a single effectively ferromagnetic 
layer with total thickness $h = 11X -7$\AA.
Hence, $\nu (X) = 9/(11X-7)$ varies from 0.072 ($X$ =12)
to 0.6 ($X$ = 2). 
The stripe periods, $D$ = 260 nm for
$X = 8$ \cite{Hellwig05}, are much larger than the
layer thicknesses  $h$ ranging from 1.5 to 12.5~nm.
Thus, the approach Eq. (\ref{energy3}) can be applied to
describe these multidomain states.
In particular, the critical line $h_t(N)$ of the 
first order transition  between the
homogeneous antiferromagnetic state and F stripes
is derived from the following equation
\begin{eqnarray}
 \pi^2 \varkappa(1-1/N)/2 = 
\exp( - \pi l/(h N) -g_N \nu +1/2)
\label{transition}
\end{eqnarray}
and plotted in the Inset of Fig.~\ref{hN}.
It should be stressed that
despite the fact that energy (\ref{energy3})
has been reduced to the same functional form 
as the energy for an isolated layer, 
these two model have 
different physical properties.
Namely the function $g_N(\nu)$ in
Eq.~(\ref{energy3}) describes
the influence of the internal ``charges''
on the equilibrium states of stripes.
The transition line $h_{t0} (N)$ between the antiferromagnetic
homogeneous state and F stripes for
an isolated layer ($g_N(\nu)$ =0)
is plotted for comparison 
in the Inset of Fig.~\ref{hN}.
The difference between the two line
$(h_{t0}-h_{t})/h_{t0} = sN/(\pi l)$
depends on the  ratio $s/l$ and
increases with increasing $N$.

In conclusion, we have investigated a continuum approach 
for the stripe multidomains that arise as regular 
equilibrium depolarization structure of coupled multilayers.
The zero-field phase diagram displays transitions
between stripe states where adjacent layers are parallel
due to depolarization effects
and other phases where the layers are antiparallel 
under influence of short-range couplings. 
In this AF case, it is possible to stabilize monodomain
states in thin multilayer systems. 
Furthermore, antiferro stripe phases can become stable 
in systems with sufficiently strong interlayer coupling. 
The transitions between these phases are first-order. 
In fact, within this model, the ferro stripe structure 
is metastable in the whole parameter range of the model.
Therefore, the different multidomain phases can coexist 
in multilayer systems and are transformed into each other
by domain nucleation and growth processes.
Hence, severe hysteresis effects can arise 
in such multilayer systems subject to specific 
coercivity mechanisms.
In particular the AF monodomain state can arise in the form of 
two domains that transform into each other 
by a global reversal of the magnetization structure.
These domains can coexist out of equilibrium as remnant 
of an antiferro stripe state.

%
\textit{Acknowledgments}. 
We thank I. Dragunov, O. Hellwig, N. Kiselev,
J. McCord, V. Neu, and R. Sch\"afer
for discussions.
A.N.B.\ thanks H.\ Eschrig for support and hospitality at the IFW Dresden. 
This work was supported by Deutsche Forschungsgemeinschaft 
through project RO 2238/6.
%
%



\begin{thebibliography} {99}


 


\bibitem{Kittel}
C. Kittel,
Phys. Rev. \textbf{70}, 965 (1946).

\bibitem{Hubert98}
A. Hubert, R. Sch{\"{a}}fer R., 
{\textit{Magnetic Domains}}
(Springer-Verlag, Berlin, 1998);
%
V. G. Bar'yakhtar, et al.
Usp. Fiz. Nauk. {\textbf{156}}, 47 (1988),
[Sov. Phys. Usp. {\textbf{31}}, 810 (1988)].


\bibitem{Scott06}
M. Dawber, K. M. Rabe, J. F. Scott,
Rev.\ Mod.\ Phys.\ \textbf{77}, 1083 (2005),
J. F. Scott,
Nanoferroelectrics: statics and dynamics
J. Phys.:Cond. Mat. \textbf{18}, R361 (2006).

%
\bibitem{Streiffer02}
S.K. Streiffer, 
J. A. Eastman, D. D. Fong, C.Thompson, A. Munkholm, 
M.V. Ramana Murty, O. Auciello, G. R. Bai, G. B. Stephenson,
Phys.\ Rev.\ Lett.\ \textbf{89}, 067601 (2002);
%
D. D. Fong, 
G. B. Stephenson, S. K. Streiffer, 
J. A.  Eastman, O.Auciello, P. H. Fuoss, C. Thompson,
Science \textbf{304}, 1650 (2005);
%
A. Schilling, T. B. Adams, R. M. Bowman, J. M. Gregg, 
G. Catalan, J. F. Scott,
Phys.\ Rev.\ B\ \textbf{47}, 024115 (2006).
%
%
%
\bibitem{Bratkovsky01}
A. M. Bratkovsky, A. P. Levanyuk,
Phys.\ Rev.\ Lett.\ \textbf{86}, 3642 (2001);
Phys.\ Rev.\ B\ \textbf{65}, 094102 (2002).

\bibitem{Fullerton03}
E. E. Fullerton et al.,
IEEE Trans. Magn. {\textbf{39}}, 639 (2003);
J. $\dot{A}$kerman,
Science \textbf{308}, 508 (2005).


\bibitem{MalekKambersky}
Z. M\'{a}lek, V. Kambersk\'{y},
Czech. J. Phys. \textbf{8}, 416 (1958).

\bibitem{KooyEnz60}
C.Kooy, U. Enz,
Philips Res. Repts. \textbf{15}, 7 (1960).

\bibitem{Suna85}
A. Suna, 
J.\ Appl.\ Phys.\ \textbf{59}, 313 (1985);
H. J. G. Draaisma, W. J. M. de Jonge, 
J.\ Appl.\ Phys.\ \textbf{62}, 3318 (1987).
%
%
%
\bibitem{Grunberg86}
P. Gr\"unberg, R. Schreiber, Y. Pang, M. B. Brodsky, and H. Sowers,
Phys.\ Rev.\ Lett.\ textbf{57}, 2442 (1986).
%
%
\bibitem{Lee04}
H.N. Lee, 
H.M. Christen, M.F. Chisholm,
C.M. Rouleau, and D.H. Lowndes,
Nature \textbf{433}, 395 (2005);
%
J.  Sigman, 
D. P. Norton, H. M. Christen, P.H. Fleming, L.A. Boatner,
Phys.\ Rev.\ Lett.\ \textbf{88}, 097601 (2002);
H. M. Christen, 
E. D. Specht, S. S. Silliman, K. S. Harshavardhan,
Phys.\ Rev. B.\ {\textbf{68}}, 020101R (2003).


\bibitem{Gehanno97}
V. Gehanno, Y. Samson, A. Marty, B. Gilles, A. Chamberod,
J.\ Magn.\ Magn.\ Mater. \ \textbf{172}, 26 (1997).
%
%

\bibitem{Hamada02}
S. Hamada, K. Himi, T. Okuno, K. Takanashi,
J.\ Magn.\ Magn.\ Mater. \ \textbf{240}, 539 (2002).

\bibitem{Hellwig03}
O. Hellwig, et al.
Nature Mater. \textbf{2}, 112 (2003);
%
O. Hellwig, A. Berger, E. E. Fullerton,
Phys.\ Rev.\ Lett.\ \textbf{91}, 197203  (2003);


\bibitem{Hellwig05}
J. Magn. Magn. Mater. \textbf{290-291}, 1 (2005).
%
%
\bibitem{Itoh03}
H. Itoh, et al. 
J. Magn. Magn. Mater. \textbf{257} 184 (2003);
%
Z.Y. Liu, S. Adenwalla,
Phys.\ Rev.\ Lett.\ \textbf{91}, 037207 (2003).
%
\bibitem{PRB04}
U. K. R{\"o}{\ss}ler, A. N. Bogdanov,  
Phys.\ Rev. B.\ {\textbf{69}}, 094405 (2004);
%
J. Magn. Magn. Mater. {\bf 269}, L287 (2004);
%
Phys.\ Rev. B.\ {\textbf{69}}, 184420 (2004).
%
%
\bibitem{Bratkovsky00}
A. M. Bratkovsky, A. P. Levanyuk,
Phys.\ Rev.\ Lett.\ \textbf{84}, 3177 (2000);
Phys.\ Rev. B \textbf{63}, 132103 (2001);
%
V. A. Stephanovich, et al.
Phys.\ Rev.\ Lett.\ \textbf{94}, 047601 (2005).
%
%

%
\bibitem{Lagerwall}
S. T. Lagerwall. Ferroelectric and Antiferroelectric 
Liquid Crystals (Wiley-VHC, Weinheim, 1999),
 L.U.Ruibo, et al.
Jpn. J. Appl. Phys. \textbf{42}, 1628 (2003).
%
\bibitem{Sayar03}
M. Sayar, M. Olivera de la Cruz, S.I. Stupp,
Europhys. Lett. \textbf{61}, 334 (2003).
%
\bibitem{Ruzette05}
A.-V. Ruzette, L. Leibler,
Nature Mater. \textbf{4} 19 (2005).
%
\bibitem{Gillijns05}
W. Gillijns,
A. Y. Aladyshkin, M. Lange, M. J. Van Bael, V.V. Moshchalkov,
Phys.\ Rev.\ Lett.\ \textbf{95}, 227003 (2005);
%
V. Jeudy, C. Gourdon, and T. Okada,
Phys.\ Rev.\ Lett.\ \textbf{92}, 147001 (2004).
%
%
\bibitem{Dong04}
J. W. Dong, 
J. Q. Xie, J. Lu, C. Adelmann, C. J. Palmstr{\o}m,
J. Cui, Q. Pan, T. W. Shield, R. D. James,
J.\ Appl.\ Phys.\ \textbf{95}, 2593 (2004);
%
A. N. Bogdanov, 
A. DeSimone, S. M\"uller, U.K. R\"o\ss ler,
J.\ Magn.\ Magn.\ Mater. \ \textbf{261}, 204 (2003).

\bibitem{Choe99}
S.-B. Choe, S.-C. Shin,
Phys.\ Rev.\ B \textbf{59}, 142 (1999).



%
\bibitem{Choe99}
S.-B. Choe, S.-C. Shin,
Phys.\ Rev.\ B \textbf{59}, 142 (1999).
%
\bibitem{Labrune01}
M. Labrune, A. Thiaville,
Eur.\ Phys.\ J.\ B \textbf{23}, 17 (2001).
%
\bibitem{FTT80}
 A. N. Bogdanov, D. A. Yablonskii.
Fiz. Tverd. Tela {\textbf{22}}, 680 (1980),
[Sov. Phys. Solid State {\textbf{22}}, 399 (1980)].
%
\bibitem{PRBstripes}
N.S. Kiselev, I.E. Dragunov, U. K. R{\"o}{\ss}ler, A. N. Bogdanov,  
to be published.
%
\bibitem{Kiselev}
The particular behaviour of $D(h)$
was first obtained for a two-layer system 
in Ref.~[\onlinecite{PRBstripes}].

\end{thebibliography}
\end{document}